\documentclass{Interspeech2024}
\usepackage{booktabs}
\usepackage{xfrac}
\usepackage{xcolor}
\usepackage{tablefootnote}
\usepackage{multirow}
\usepackage{cleveref}

\interspeechcameraready

\definecolor{xiaomi_gray}{HTML}{A9A9A9}
\definecolor{catpuccin_red}{HTML}{f38ba8}
\definecolor{catpuccin_blue}{HTML}{89b4fa}
\title{Scaling up masked audio encoder learning for general audio classification}

\name{Heinrich}{Dinkel}
\name{Zhiyong}{Yan}
\name{Yongqing}{Wang}
\name{Junbo}{Zhang}
\name{Yujun}{Wang}
\name{Bin}{Wang}

\address{
  AI Lab, Xiaomi Corporation, China}
\email{\{dinkelheinrich,yanzhiyong, wangyongqing3, zhangjunbo1, wangyujun, wangbin11\}@xiaomi.com}

\keywords{Audio classification, General audio feature, Transformer, Masked auto encoder}

\begin{document}

\maketitle

\begin{abstract}
Despite progress in audio classification, a generalization gap remains between speech and other sound domains, such as environmental sounds and music.
Models trained for speech tasks often fail to perform well on environmental or musical audio tasks, and vice versa.
While self-supervised (SSL) audio representations offer an alternative, there has been limited exploration of scaling both model and dataset sizes for SSL-based general audio classification.
We introduce Dasheng, a simple SSL audio encoder, based on the efficient masked autoencoder framework. 
Trained with 1.2 billion parameters on 272,356 hours of diverse audio, Dasheng obtains significant performance gains on the HEAR benchmark.
It outperforms previous works on CREMA-D, LibriCount, Speech Commands, VoxLingua, and competes well in music and environment classification.
Dasheng features inherently contain rich speech, music, and environmental information, as shown in nearest-neighbor classification experiments.

\end{abstract}

\section{Introduction}

In recent years, machine learning applications, especially in text and vision processing, have experienced significant advancements, primarily driven by the pretraining of large models on extensive datasets. 
For instance, in the field of vision, ImageNet is widely acknowledged as the standard pretraining dataset. 
Models trained on ImageNet in a supervised manner find widespread applicability in various classification and separation tasks within the vision domain.
However, this transfer capability of vision models across tasks is not currently observed in the domain of audio classification.
For example, in~\cite{DinkelYWZW22} authors have shown that supervised pretraining on AudioSet (similar to ImageNet in audio), yields benefits solely for sound classification tasks. 
Conversely, other tasks such as language classification, speaker recognition, and intent classification were found to adversely impact performance with supervised AudioSet pretraining.

Research aimed at developing comprehensive backbones for general audio classification has also been recently accelerated by benchmarks such as the Holistic evaluation of Audio Embeddings (HEAR)~\cite{turian2022hear}.
Results on the HEAR benchmark suggest that self-supervised models (SSLs) are likely to be more potent than their supervised counterparts when it comes to general performance on a variety of audio tasks.
\begin{figure}[tb]
    \centering
    \includegraphics[width=0.97\linewidth]{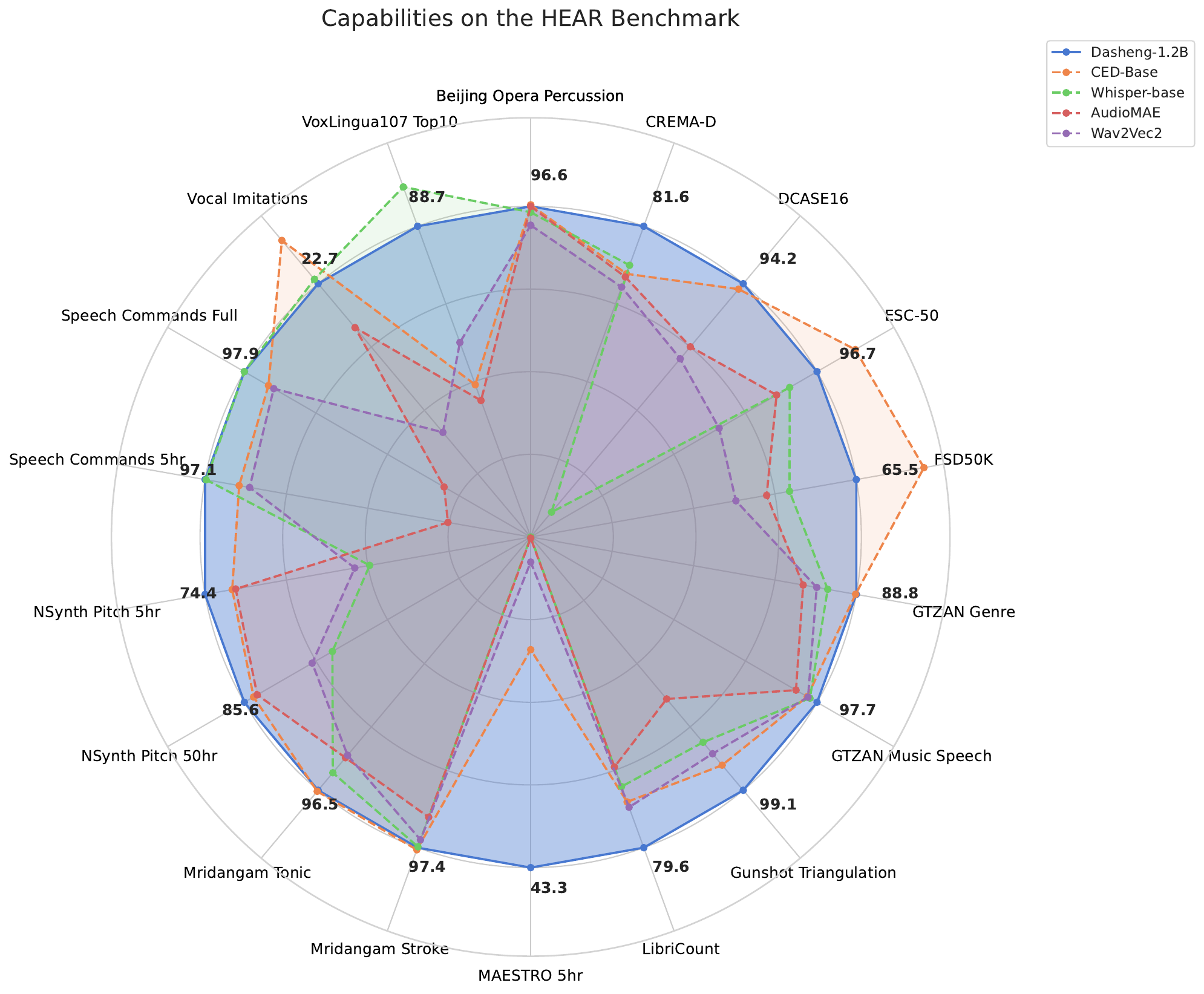}
    \caption{Graph showcasing Dasheng's capability on the HEAR benchmark, compared to expert models CED-Base (Environment, Music) and Whisper-base (Speech), as well as baselines AudioMAE and Wav2Vec2. Best viewed in color.
    }
    \label{fig:hear_capabilities}
\end{figure}
Currently, a variety of self-supervised training paradigms exist.
In the next token prediction approaches such as Wav2Vec2~\cite{baevski2020wav2vec}, the model is tasked to predict future high-level audio tokens given a context of past tokens.
Further, in masked token prediction~\cite{he2022masked,chen2022beats,niizumi2022masked}, segments of data are randomly erased (zeroed) and the model is tasked to predict the masked tokens from a provided context.
Finally, in bootstrap your own latent (BYOL)~\cite{grill2020bootstrap,niizumi2023byol-a}, a student model is tasked to predict hidden representations of a teacher model.

However, to the best of our knowledge, there has been little emphasis on scaling up the SSL representation encoder models (beyond 300M~\cite{huang2022masked}) by increasing the parameter size.
One of the potential reasons is the limited amount of publicly available large-scale general audio datasets.
In recent years, the 5000-hour-long AudioSet~\cite{gemmeke2017audio} has been extensively utilized for most general representation learning tasks, while larger datasets such as ACAV100M~\cite{lee2021acav100m} have not been explored before.
Another hindrance is the computational overhead stemming from enlarging the model parameter size, requiring large clusters of graphics processing units (GPUs).

In our point of view, the most effective SSL approach for scalable pretraining is the use of masked autoencoders (MAE)~\cite{he2022masked}.
These types of models deviate from conventional masked learning by removing masked-out data, thereby substantially decreasing computational overhead and facilitating the scalability of models.
Even though MAEs have been proposed~\cite{huang2022masked} and improved for the audio domain~\cite{niizumi2022masked,niizumi2023masked}, previous works mainly used \SI{86}{M} parameter models with the aforementioned AudioSet training data.
Thus, this work proposes a general audio classification back-bone, named \textbf{D}eep \textbf{A}udio-\textbf{S}ignal \textbf{H}olistic \textbf{E}mbeddi\textbf{ng}s (Dasheng).
Dasheng is an MAE-style pre-trained encoder model, that has been scaled to 1.2 billion parameters and trained on 272,356 hours of publicly available data, showing impressive performance across a variety of audio classification tasks (\Cref{fig:hear_capabilities}).
While our primary focus lies in exploring classification performance, it is important to note that MAE's can also be used for audio generation~\cite{liu2023audioldm} and superresolution~\cite{fre_painter2024} tasks, boasting state-of-the-art (SOTA) performance in both domains.


\section{Approach}
\label{section:approach}

\begin{figure}
    \centering
    \includegraphics[width=0.99\linewidth]{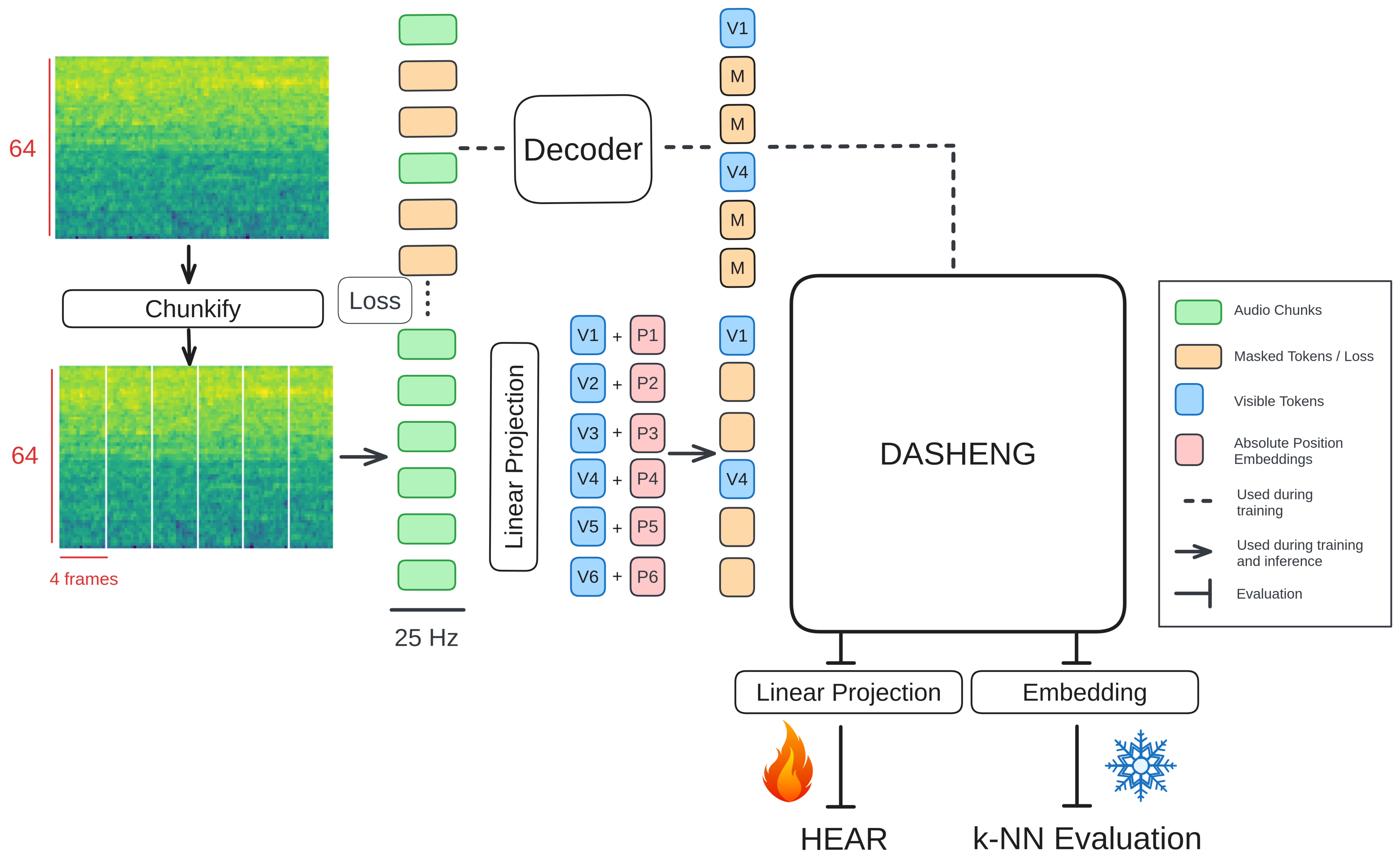}
    \caption{The Dasheng training framework. Four consecutive Mel-spectrogram frames are ``chunkified'' into a single token. 
    Following a linear transformation and the addition of a positional embedding, 75\% of these chunked representations are discarded.
    The resulting tokens $\mathbf{V}$ are then fed into Dasheng, which extracts high-dimensional embeddings. 
    During training, these embeddings are further fed into a small decoder responsible for predicting those chunks that were initially excluded.}
    \label{fig:framework}
\end{figure}

Dasheng is based on the MAE framework, which is composed of a transformer-based asymmetric encoder-decoder, where only the decoder operates on the entire input sequence, enabling efficient encoder training.
Given a Mel-spectrogram of size $\mathbf{X}_{\text{mel}} \in \mathbb{R}^{\mathtt{T} \times \mathtt{F}}$, where $T$ represents the number of frames and $F$ the number of filterbanks, we first proceed to split the time-axis into equally sized chunks and project each chunk by a linear transformation to a specified dimension as: $\mathbf{X}_{\text{mel}} \mapsto \mathbf{V} \in \mathbb{R}^{N \times D}$, where $N$ represents the number of chunks or ``tokens'' and $D$ is the model's embedding dimension.
We further add absolute learnable positional embeddings $\mathbf{P} \in \mathbb{R}^{N \times D}$ to $\mathbf{V}$.
Then, we mask $\mathbf{M} \in \{0,1\}$ and discard 75\% of the tokens and feed the unmasked tokens $\mathbf{X}_{\text{encoder}} = \mathbf{V\odot (\mathbf{1}-\mathbf{M})}$ to an encoder, which predicts embeddings $\text{Encoder}(\mathbf{X}_{\text{encoder}}) \mapsto \mathbf{E} \in \mathbb{R}^{N_{\text{unmask}} \times D}$.
Then, we obtain $\mathbf{X}_{\text{decoder}}$ by appending a learnable mask token to $\mathbf{E}$, for each position that has been masked.
The decoder predicts $\text{dec}(\mathbf{X}_{\text{decoder}}) \mapsto \mathbf{\hat{X}}_{\text{mel}}$.
Finally, we calculate the normalized mean square error (MSE) loss between the predicted masked tokens and the ``chunkified'' ground truth~\cite{he2022masked} spectrogram.
The entire training framework is depicted in \Cref{fig:framework}.
After training, we freeze all parameters of the encoder and evaluate its embeddings on a variety of downstream tasks.

Dasheng differentiates itself from previous MAE-based works~\cite{huang2022masked}, through several key differences.
Notably, Dasheng incorporates learnable absolute positional embeddings, produces frame-level embeddings at a higher frequency of 25 Hz (compared to the 6.25 Hz used in prior approaches), and operates on consecutive chunks of Mel-spectrogram frames rather than adopting the more conventional time-frequency ``patch'' representation seen in earlier methods.



\section{Experiments}
\label{section:experiments}

\subsection{Datasets}

\subsubsection{Training datasets}
To enable generalization for speech, music, and environmental sounds, our work utilizes general audio datasets being AudioSet~\cite{gemmeke2017audio}, ACAV100M~\cite{lee2021acav100m}, and VGGSound~\cite{chen2020vggsound}.
AudioSet, VGGSound, and ACAV100M encompass audio clips sourced from YouTube videos, offering a high diversity of general audio. 
Each audio clip in VGGSound and AudioSet is identified solely by the presence of sound/visual event tags.
In contrast, ACAV100M comprises 100 million videos that have been filtered from a larger superset by strong audio-visual correlation, i.e., sounds and visual cues are likely synchronized.
Due to partial unavailability and difficulties acquiring the above-mentioned datasets, we provide in-depth information about our downloaded dataset in \Cref{tab:datasets}.
Only the audio contained in each video is used in this work.
We also added MTG-Jamendo~\cite{bogdanov2019mtg} as a source to further enhance performance for music tasks.
During training, we discard any labels present in MTG-Jamendo, VGGSound, and AudioSet, and for each epoch, we evaluate on the held-out test subset of VGGSound in regards to MSE performance.
\begin{table}[tb]
    \centering
    \caption{Training datasets used in this work.}
    \label{tab:datasets}
    \begin{tabular}{l|rrr}
    \toprule
        Dataset & \# Samples & Duration (h) & Type \\
        \midrule
        ACAV100M & 94,934,272 &  263,000 & General \\
        AudioSet & 1,904,747 & 5,100 & General \\
        VGGSound & 176,819 & 488 & General \\    
        MTG-Jamendo & 55,701 &  3,768 & Music \\
        \midrule
        All & 97,071,539 & 272,356 & General \\
        \bottomrule
    \end{tabular}
\end{table}

\begin{table*}[htp]
    \centering
    \caption{Main results on the HEAR benchmark dataset across 18 tasks. ``HEAR SOTA'' signifies the leading result on the HEAR leaderboard, primarily achieved through individual models. Boldface indicates surpassing the best HEAR model and higher is better for all values.
 Models marked with $^\star$ have been evaluated from a publicly available checkpoint and results in {\color{xiaomi_gray}{gray}} have been trained on the respective dataset. Models displayed in {\color{red}red} have been trained in supervised fashion and {\color{blue} blue} represent SSL-based approaches. }
    
    \label{tab:results}
    \resizebox{\textwidth}{!}{%

\begin{tabular}{l|cccccccccccccccccc}
Model & BJ & CD & D16 & E50 & F50k & GZ-Gen & GZ-M/S & Gun & LiCt & MST & Mri-S & Mri-T & NS-50 & NS-5 & SPC-5 & SPC-F & VI & VL \\
\midrule
{\color{red} CED-Base}~\cite{dinkel2023ced} & 96.6 & 69.1 & 92.2 & 96.7 & 65.5 & 88.6 & 94.4 & 89.3 & 67.9 & 14.8 & 97.4 & 96.6 & 82.8 & 68.2 & 86.9 & 89.7 & 22.7 & 38.6 \\
{\color{red} Whisper-base}$^\star$~\cite{radford2023robust} & 94.5 & 71.3 & 9.2 & 77.1 & 43.1 & 81.0 & 95.3 & 80.4 & 63.9 & 0.0 & 96.5 & 89.5 & 59.3 & 36.8 & 96.8 & 97.9 & 19.7 & \color{xiaomi_gray}{88.7} \\
\midrule
{\color{blue}BYOL-S}~\cite{turian2022hear} & 95.3 & 65.7 & 64.2 & 80.5 & 50.9 & 83.7 & 93.9 & 85.7 & 78.5 & 0.8 & 97.3 & 92.9 & 71.2 & 39.6 & 91.4 & 94.8 & 16.0 & 45.8 \\
{\color{blue}HuBERT-FUSE}~\cite{turian2022hear} & 94.9 & 75.2 & 82.6 & 74.4 & 41.3 & 79.6 & 93.6 & 92.9 & 68.3 & 16.6 & 97.4 & 90.9 & 68.8 & 38.2 & 94.7 & 95.7 & 18.5 & 71.4 \\
{\color{blue}Sony-VIT}~\cite{turian2022hear} & 92.8 & 44.1 & 66.8 & 40.1 & 0.0 & 68.1 & 89.9 & 86.6 & 48.8 & 23.9 & 88.0 & 73.0 & 79.5 & 69.2 & 47.9 & 53.1 & 6.8 & 22.4 \\
{\color{blue}Wav2Vec2}~\cite{turian2022hear} & 90.7 & 65.6 & 66.3 & 56.1 & 34.2 & 78.0 & 94.6 & 84.8 & 69.2 & 3.3 & 94.3 & 82.8 & 65.3 & 40.2 & 83.8 & 87.9 & 8.0 & 49.3 \\

{\color{blue} MSM-MAE}~\cite{niizumi2022masked} & 94.9 & 73.4 & 0.0 & 85.6 & 52.2 & 86.1 & 99.2 & 96.4 & 77.8 & 0.0 & 97.5 & 98.3 & 81.2 & - & - & 86.4 & 18.3 & 50.0 \\
{\color{blue} AudioMAE}$^\star$~\cite{huang2022masked} & 96.2 & 68.3 & 70.8 & 73.2 & 39.3 & 74.3 & 90.5 & 63.4 & 58.9 & 0.1 & 87.3 & 83.8 & 81.7 & 67.4 & 24.7 & 29.6 & 16.0 & 34.6 \\
{\color{blue} Data2Vec}$^{\star}$~\cite{baevski2022data2vec} &  85.6 & 54.2 & 43.3 & 34.9 & 17.9 & 57.3 & 96.1 & 67.0 & 55.4 & 0.2 & 86.0 & 74.0 & 23.9 & 12.8 & 91.9 & 94.2 & 8.7 & 42.9 \\
{\color{blue} Beats$_{\text{Iter1}}$}$^\star$~\cite{chen2022beats}& 95.8 & 68.1 & 43.0 & 81.9 & 51.4 & 87.0 & 98.5 & 90.5 & 74.6 & 0.0 & 96.1 & 96.0 & 82.0 & 68.6 & 88.2 & 91.5 & 13.5 & 43.8 \\
{\color{red} Beats$_{\text{Iter3+}}$}$^\star$~\cite{chen2022beats} & 96.2 & 71.3 & 39.6 & 94.5 & 63.6 & 90.0 & 96.8 & 95.2 & 76.4 & 0.1 & 96.8 & 94.7 & 81.4 & 64.2 & 83.9 & 86.7 & 18.4 & 37.8 \\
{\color{blue}ATST-Frame}~\cite{li2024self} & 95.8 & 76.7 & 95.7 & 89.0 & 55.7 & 88.3 & 100.0 & 94.3 & 78.1 & 24.4 & 97.5 & 94.1 & - & 68.6 & 92.6 & 95.1 & 22.3 & 66.9 \\
{\color{blue}ATST-Clip}~\cite{li2024self} & 95.3 & 76.0 & 93.7 & 91.2 & 59.5 & 87.7 & 99.2 & 98.8 & 78.2 & 18.9 & 97.7 & 96.7 & - & 67.8 & 93.1 & 95.5 & 18.5 & 53.9 \\
\midrule
HEAR SOTA & 97.5 & 76.7 & 95.7 & 96.7 & 65.5 & 96.7 & 100.0 & 100.0 & 78.5 & 46.9 & 97.7 & 96.7 & 90.0 & 87.8 & {\color{xiaomi_gray}97.6} & {\color{xiaomi_gray}97.8} & 22.7 & 72.2 \\
\midrule
{\color{blue} Dasheng-Base} & 93.6 & 78.7 & 93.9 & 82.9 & 51.0 & 89.2 & 99.2 & 92.9 & 76.6 & 43.9 & 96.1 & 94.9 & 83.3 & 71.8 & 95.9 & 97.1 & 16.7 & 69.9 \\
{\color{blue} Dasheng-0.6B} & 94.9 & 81.2	 & 94.4	 & 85.9 & 	53.9 &	88.6	& 97.6 & 	97.6 & \textbf{80.7} &	43.5 & 	96.6 &	96.2 &	85.8 &	74.6 &	97.0 &	97.5 & 	17.8 &	{74.7} \\
{\color{blue} Dasheng-1.2B} & 96.2 & \textbf{81.6} & 94.2 & 85.3 & 54.2 & 88.8 & 97.7 & 99.1 & 79.6 & 43.3 & 96.8 & 96.1 & 85.6 & 74.4 & 97.1 & \textbf{97.9} & 19.4 & \textbf{78.7} \\

\bottomrule
\end{tabular}

    }
\end{table*}

\subsubsection{Downstream datasets}
The study primarily evaluates its results on the HEAR benchmark~\cite{turian2022hear}.
HEAR comprises 19 tasks, broadly categorized into Speech, Music, and Environment.
The tasks within the Speech category include SpeechCommands (SPC) 5h/Full, Voxlingua (VL), LibriCount (LiCt), Vocal Imitations (VI), and CREMA-D (CD). 
In the Music category, tasks encompass Beijing Opera (BJ), GTZAN Genre (GZ-Gen), GTZAN Music/Speech (GZ-M/S), Mridangam Tonic (Mri-T), Mridangam Stroke (Mri-S), MEASTRO 5h (MST) and NSynth (NS) Pitch 5h/50h. 
Lastly, tasks related to the Environment include Beehive, DCASE16 (D16), ESC-50 (E50), and FSD50k (F50k).
Further one can divide these tasks into 17 clip-level and two frame-level (D16, MST) classification tasks. 
The HEAR benchmark trains a shallow multi-layer perceptron (MLP) classifier on top of frozen embeddings.
For further information regarding the datasets, please refer to~\cite{turian2022hear}.
Following previous works~\cite{barlow_twins}, we discard the ``Beehive'' subtask due to its overly long utterances and small sample size, leading to inconsistent results.

\subsubsection{Embedding extraction}
For all downstream tasks, we use the output of the last layer as the representative embedding, extracted at 25 Hz.
In cases where a single clip-level embedding is required, we mean-pool those frame-level embeddings.

\subsection{Setup}
\label{ssec:setup}

In regards to data processing, we resample all datasets to 16 kHz and extract 64-dimensional log-Mel spectrograms every \SI{10}{ms} with a window size of \SI{32}{ms} for clips of length \SI{10}{s}.
During training, we adopt a grouped masking strategy to address the issue that the last frame of a mel-spectrogram incorporates information from future frames, specifically, the next three frames in our case. 
To enhance overall training stability, we systematically discard at least two consecutive chunks (equivalent to 80 ms of audio) during training.
Model training uses an 8-bit AdamW optimizer~\cite{dettmers2022optimizers} with a cosine decay scheduler, starting from a learning rate of 0.0003 and a weight decay of 0.01.
The larger 1.2B model uses a learning rate of 0.0002 and no weight decay.
A training epoch involves sampling 15,000 batches, and the training duration spans 100 epochs, equivalent to 4 full data epochs on our training dataset.
A batch size of 32 per GPU is utilized across eight A100 GPUs and training takes approximately four days to finish for the 1.2B model.
We incorporate a 3-epoch warm-up for the learning rate, followed by a decay to 10\% of its maximal value over the training period.
Dasheng can process a maximum of \SI{10}{s} of audio at once. 
In downstream evaluation scenarios involving longer inputs, we segment the audio into \SI{10}{s} chunks. These segments are then forwarded through the model individually, and the resulting embeddings are concatenated.
The neural network back-end is implemented in Pytorch~\cite{PaszkePytorch} and the source code with pretrained checkpoints is publicly available\footnote{\url{https://github.com/RicherMans/Dasheng}}.

\begin{table}[t]
    \centering
    \caption{Model setups used in this study. ``Depth'' represents the number of blocks in the model, ``Embed'' refers to the embedding dimension, ``MLP'' to the dimension of each block's multi-layer perceptron and ``\#Heads'' stands for the number of independent attention mechanisms.}
    \begin{tabular}{l|rrrrr}
        \toprule
        Model &  \# Param & Depth & Embed & MLP & \#Heads \\
        \midrule
        Base & \SI{86}{M} & 12 & 768 & 3072 & 12  \\
        0.6B &  \SI{600}{M} &  32  & 1024 & 4096 & 16 \\
        1.2B & \SI{1200}{M} & 40 & 1536 & 6144 & 24 \\
        \hline
        Dec-25M & \SI{25}{M} & 8 &  512 & 2048 & 16 \\
        Dec-56M & \SI{56}{M} & 8 &  768 & 3072 & 24 \\
        
        \bottomrule
    \end{tabular}
    
    \label{tab:models}
\end{table}

To showcase the scaling capabilities of MAEs, we train three differently-sized models, seen in \Cref{tab:models}.
Dasheng-Base represents the most commonly used model parameter size in literature, to which we add a 0.6B and 1.2B parameter model.
All models share the setup with common vision transformers (ViT)~\cite{Dosovitskiy_ViT}, using pre-norm and GeLU activations.
During the training phase, we attach Dec-25M to the Base and 0.6B encoders, while the larger Dec-56M is used for the 1.2B model.

\section{Results}
\label{section:results}

\subsection{HEAR}
\label{ssection:hear}

The results regarding the HEAR benchmark can be seen in \Cref{tab:results}.
Furthermore, we present top-line results achieved through supervised pretraining approaches, specifically utilizing CED-Base~\cite{dinkel2023ced} for environment/music tasks and Whisper-base~\cite{radford2023robust} for speech. 
Additionally, we incorporate a range of current state-of-the-art SSL embeddings~\cite{li2024self,chen2022beats} along with commonly used audio representations like Wav2Vec2 and BYOL.
Dasheng 1.2B achieves impressive performance on the majority of tasks, scoring over 80 on 13 out of 18. 
Notably excelling in Emotion Recognition (CD), Language Identification (VL), and Keyword Spotting (SPC), while Dasheng 0.6B stands out in Speaker Counting (LiCt).
For Emotion Recognition, Dasheng 1.2B significantly outperforms previous attempts (76.7 $\rightarrow$ 81.6). 
Particularly noteworthy is its 78.7\% accuracy in Language Identification, surpassing Wav2Vec2, trained on 100k hours of multilingual speech.
It's worth highlighting that the best-reported result on VL (Whisper, 88.7\%) involved the use of \SI{650}{k} hours of supervised multilingual speech training data.
Another notable result is in pitch estimation on NS-50, where the proposed models can achieve a score of up to 85. 
This is slightly below the current SOTA score of 90, achieved by a dedicated pitch estimator.
In frame-level classification tasks, the proposed models excel, achieving results of 94.4 in D16 and 43.9 in MST, just below the SOTA performances of 95.7 and 46.9, respectively. 
For a visual representation of capabilities across HEAR, refer to \Cref{fig:hear_capabilities}.

\subsection{Cross-domain capabilities}

In this section, we summarize our findings from \Cref{tab:results} into three audio categories: Environment (Env), Speech, and Music, by averaging scores for each subtask, and results are presented in \Cref{tab:summary_results_hear}.
Notably, in environment classification, Dasheng is surpassed by CED-Base, which, however, performs poorly in speech classification. 
Dasheng excels in speech-related tasks, surpassing Whisper-Base and Wav2Vec2.
Lastly, likely due to the amount of music-related training data, all proposed models significantly outperform previous works in music-related tasks.
On average, the proposed models outperform previous works, demonstrating versatility across the audio domain.

\begin{table}[htbp]
    \centering
    \caption{Performance regarding Environment (Env), Speech, and Music tasks within the HEAR benchmark. The best results per category are highlighted in bold and higher is better. }
      \label{tab:summary_results_hear}
    \begin{tabular}{l|rrr|r}
    \toprule
         Model & Env & Speech & Music & Average  \\
\midrule
CED-Base & \textbf{85.90} &	62.47 &	79.91 &	76.09 \\
Whisper-Base & 52.45&	73.06	& 69.10 &	64.87 \\
\midrule
Wav2Vec2 & 60.35	 & 60.63 & 	68.65 & 63.21 \\
MSM-MAE &  78.07	& 61.18 &	69.65 &	69.63 \\
AudioMAE & 61.66&	38.68&	72.66&	57.67 \\
Data2Vec & 40.77 & 57.87 & 54.48 & 51.04 \\
Beats$_{\text{Iter1}}$ & 66.68	&63.28 &	78.00 &	69.32 \\
Beats$_{\text{Iter3+}}$ & 73.23 &	62.40 &	77.52 &	71.05 \\
ATST-Frame & 83.68 &	71.95 &	71.09	& 75.57 \\
ATST-Clip & {85.80}	& 69.20 &	70.41 &	75.14 \\
            \midrule
        Dasheng-Base & 80.15	 & 72.48 & 	84.01	& 78.88\\
         Dasheng-0.6B & 82.38 &	74.89	 & 84.04 &	80.44 \\
         Dasheng-1.2B & 83.20 & \textbf{75.71} & \textbf{84.86} & \textbf{81.25} \\
         \bottomrule
    \end{tabular}

\end{table}

\vspace{-4mm}
\subsection{Classification with k-nearest neighbors}
\label{ssec:knn}

In this section, we are interested in further exploring the performance of the embeddings without parameterized, supervised fine-tuning, and thus, we perform simple k-nearest neighbor (k-NN) classification on nine tasks.
We specifically use the Fluent Speech Commands (FSC)~\cite{lugosch2019speech}, UrbanSound8k (US8k)~\cite{salamon2014dataset}, NSynth Instrument (NS$_{\text{Inst}}$)~\cite{engel2017neural}, VoxCeleb1~\cite{nagrani2017voxceleb}, RAVDESS-Speech~\cite{livingstone2012ravdess}, FSDKaggle2018~\cite{fonseca2017freesound} (FSDK18), Speechcommands 1 and 2 (10/35 class) and ESC-50~\cite{piczak2015esc}.
All tasks are assessed at the clip-level by mean-pooling all frame-level representations, employing a consistent setting with $k = 10$ for evaluation, and accuracy serves as the primary metric.
Performance for RAVDESS, US8k, and ESC-50 is assessed through k-fold cross-validation, while other tasks use the held-out test set.
For comparison, we include an AudioMAE~\cite{huang2022masked} baseline. 
The results in \Cref{tab:knn_eval} indicate that Dasheng embeddings are inherently potent representations for general audio classification tasks.
Notably, for both keyword spotting tasks (SPC1/2), Dasheng 1.2B achieves an impressive accuracy of 95\% and 90.7\%, respectively, significantly outperforming the AudioMAE baseline.
Also, instrument classification on NS achieves an accuracy of 71.2\%, indicating strong capabilities in music classification. 
Surprisingly, results for speaker recognition (VoxCeleb1) suggest that speaker information is present in Dasheng, achieving an accuracy of up to 39.4\%, significantly outperforming the baseline.
This surprising result prompted us to run a linear evaluation on VoxCeleb1, achieving 82.5\%, 89.4\%, and 92.5\% for the Base, 0.6B, 1.2B models, respectively.
In the future, we aim to conduct further experiments with Dasheng, specifically focusing on speech recognition and speaker identification.

\begin{table}[htbp]
    \centering
    \caption{Evaluation using a k-NN classifier where values represent accuracy on each task's test-set and $k=10$. Best in bold. }
    \label{tab:knn_eval}
\begin{tabular}{ll|c|rrr}
\toprule
\multirow{2}{*}{Domain} & \multirow{2}{*}{Task}& \multirow{2}{*}{AudioMAE} & \multicolumn{3}{c}{Dasheng} \\
 &   &  & Base & 0.6B & 1.2B \\
\hline
\multirow{3}{*}{{Env}}  & ESC50  & 53.1 & 61.9 & 66.6 & \textbf{68.6} \\ 
 & FSDK18  & 43.4 & 70.3 & \textbf{72.1} & \textbf{72.1} \\
& US8k  & 58.2 & 73.9 & 75.9 & \textbf{77.7} \\
\hline
\multirow{1}{*}{{Music}}  & NS$_{\text{Inst}}$  & 67.2 & 70.0 & 70.9 & \textbf{71.2} \\ %
\hline
\multirow{5}{*}{{Speech}}  & SPC1  & 56.9 & 93.6 & 93.4 & \textbf{95.9} \\
& SPC2  & 5.9 & 86.0 & 87.3 & \textbf{90.9} \\
& VoxCeleb1  & 2.9 & 34.2 & 37.8 & \textbf{39.4} \\
& RAVDESS  & 28.7 & 58.1 & 61.8 & \textbf{61.9} \\
& FSC  & 7.6 & 52.3 & 57.6 & \textbf{62.4} \\


\bottomrule
\end{tabular}
    
\end{table}

\subsection{Impact of scaling to performance}

Here we evaluate the impact of scaling dataset and model size in regards to performance.
In this experiment, we conduct training on AudioSet for 30 epochs using the settings outlined in \Cref{ssec:setup}. 
The outcomes are presented in \Cref{tab:dataset_size_impact}. 
As we can see, increasing model size consistently improves performance when training on AS.
Nevertheless, jointly increasing both dataset size and model size results in significantly enhanced results, 
improving average performance by 6.37, 8.69, and 8.45 points for the Base, 0.6B, and 1.2B models, respectively.

\begin{table}[htbp]
    \centering
    \caption{Impact of training data size on performance. Results represent Average HEAR scores across three domains. ``AS'' represents AudioSet and ``Ours'' uses data described in \Cref{tab:datasets}.}
    \label{tab:dataset_size_impact}
\begin{tabular}{r|ll|r}
\toprule
\multirow{2}{*}{Model} & \multicolumn{2}{c|}{Training data} &   \multirow{2}{*}{Difference} \\
 & AS & Ours & \\
\midrule
Base & 70.43 &  78.88 & +8.45 \\
0.6B & 71.75 & 80.44 & +8.69 \\
1.2B & 74.87 & 81.25 & +6.37 \\
\bottomrule
\end{tabular}
    
\end{table}

\vspace{-3mm}
\section{Conclusion}
\label{section:conclusion}

We introduced Dasheng, a general model for audio classification tasks.
Dasheng is based on the efficient MAE framework, which made training a 1.2 billion parameter model on 272,356 hours of data with limited access to large GPU clusters, feasible.
MLP evaluation results on the HEAR benchmark show strong performance across 18 tasks, while also outperforming previous attempts on four tasks.
Notably, Dasheng achieves excellent performance in keyword spotting, language identification, speaker counting, and emotion classification while at the same time being capable of music note and genre classification and competitive in environment sound event classification.
Further k-NN evaluation reveals that Dasheng features can directly be used without parameterization for classification tasks across a variety of downstream tasks.
Most importantly, this paper provides empirical evidence that large-scale pretraining for audio representations using the MAE framework results in substantial performance improvements.



\bibliographystyle{IEEEtran}
\bibliography{mybib}

\end{document}